\documentclass[12pt]{article}
\usepackage{amsmath,amssymb,theorem,cite,epsfig,url,psfrag,eepic,mathtools,amsmath}
\advance \topmargin by -\headheight
\advance \topmargin by -\headsep     
\evensidemargin \oddsidemargin
\def\Maketitle{{\def\newpage{}\maketitle}}
\makeatletter
\def\Appendix{\appendix
  \def\@seccntformat##1{Appendix~\csname the##1\endcsname.~~}}
\makeatother
\makeatletter
\@addtoreset{equation}{section}

\makeatother

\def\XXint#1#2#3{{\setbox0=\hbox{$#1{#2#3}{\int}$}
\vcenter{\hbox{$#2#3$}}\kern-.5\wd0}}

\begin{document}
\rightline{RUNHETC-2014-18}
\title{\textbf{On spectrum of ILW hierarchy\\ in conformal field theory II: coset CFT's}\vspace*{.3cm}}
\date{}
\author{M.~N.~Alfimov$^{1,2,3,4}$ and A.~V.~Litvinov$^{5,6}$\\[\medskipamount]
$^1$~\parbox[t]{0.85\textwidth}{\normalsize\it\raggedright
LPT, Ecole Normale Superieure, 75005 Paris, France}\vspace*{2pt}\\[\medskipamount]
$^2$~\parbox[t]{0.85\textwidth}{\normalsize\it\raggedright
Insitut de Physique Theorique, CEA Saclay, 91191 Gif-sur-Yvette Cedex, France}\vspace*{2pt}\\[\medskipamount]
$^3$~\parbox[t]{0.85\textwidth}{\normalsize\it\raggedright
P.N. Lebedev Physical Institute,  119991 Moscow, Russia}\vspace*{2pt}\\[\medskipamount]
$^{4}$~\parbox[t]{0.85\textwidth}{\normalsize\it\raggedright
Moscow Institute of Physics and Technology, 141700 Dolgoprudny, Russia}\vspace*{2pt}\\[\medskipamount]
$^5$~\parbox[t]{0.85\textwidth}{\normalsize\it\raggedright
Landau Institute for Theoretical Physics,
142432 Chernogolovka, Russia}\vspace*{2pt}\\[\medskipamount]
$^6$~\parbox[t]{0.85\textwidth}{\normalsize\it\raggedright
NHETC, Department of Physics and Astronomy, Rutgers University,\\ Piscataway, NJ 08855-0849, USA}}
\Maketitle
\begin{abstract}
We study  integrable structure of the coset conformal field theory and define the system of Integrals of Motion which depends on external parameters. This system can be viewed as a quantization of the ILW type hierarchy. We propose a set of Bethe anzatz equations for its spectrum.
\end{abstract}
\section{Introduction}\label{intro}
This paper is a direct continuation of \cite{Litvinov:2013zda} where the set of Bethe anzatz equations for the spectrum of Integrals of Motion (IM) in Conformal Field Theory (CFT) was proposed. Here we consider more general class of CFT's  defined by GKO coset construction \cite{Goddard:1984vk}
\begin{equation}\label{GKO}
 \mathcal{M}(r,p,n)=\frac{\widehat{\mathfrak{sl}}(r)_{p}\times\widehat{\mathfrak{sl}}(r)_{n-p}}{\widehat{\mathfrak{sl}}(r)_{n}}.
\end{equation}
This set of CFT's is unitary for non-negative integer values of the parameters $p$ and $n-p$ and includes the $W$ unitary minimal models $\mathcal{WA}_{r}(k)=\mathcal{M}(r,1,k-r)$ as a particular case.

It is well known \cite{Ahn:1990gn} that among  primary fields of the model $\mathcal{M}(r,p,n)$ there is a special one which defines an integrable perturbation. It has  conformal dimension $\Delta=\frac{n}{n+r}$ and corresponds by GKO construction to the  branching of the product of two vacuum representations into adjoint one. This operator is a complete analog of $\Phi_{1,3}$ operator in minimal models. According to \cite{Zamolodchikov:1987jf} such an integrable perturbation defines an infinite set of commuting operators called local Integrals of Motion. In this paper we study the problem of computation of their common spectrum.

The coset CFT's \eqref{GKO} admit some chiral algebra description. Typically, it involves currents of fractional spins with non-abelian braiding which makes it difficult to analyze such models.  In some cases the description simplifies and  the chiral algebra reduces to known algebras.  For example we get $\mathrm{W}_{r}$ algebra \cite{Zamolodchikov:1985wn,Fateev:1987vh,Fateev:1987zh} for $p=1$, $\textsf{NSR}$ algebra for $r=p=2$ and spin-$\frac{4}{3}$ parafermion algebra \cite{Fateev:1985ig} for $r=2$, $p=4$. For us it will be important that the chiral algebra description of the coset models \eqref{GKO} exists and  depends smoothly on the parameter $n$ which we parameterize as
\begin{equation}
   n=\frac{p}{1+b^{2}}-r.
\end{equation}
The new parameter $b$ will be treated below as continuous and we denote the corresponding chiral algebra as $\mathcal{A}(r,p)$. In new notations its central charge can be written as
\begin{equation}\label{central-charge}
   c=\frac{p(r^{2}-1)}{p+r}+\frac{r(r^{2}-1)}{p}Q^{2},\quad\text{where}\quad Q=b+\frac{1}{b}.
\end{equation}
Sometimes it is convenient to think about the algebra  $\mathcal{A}(r,p)$ as a symmetry algebra of parafermionic Toda theory \cite{LeClair:1992xi}. This theory can be realized as a coupled theory of $(r-1)$-component bosonic field $\varphi$ and $\widehat{\mathfrak{sl}}(r)_{p}/\widehat{\mathfrak{u}}(1)^{r-1}$ parafermions  \cite{Gepner:1987sm}. The Lagrangian can be schematically written as follows
\begin{equation}\label{Lagrangian}
   \mathcal{L}=\mathcal{L}_{\textrm{PF}}+\frac{p}{8\pi}\left(\partial_{a}\varphi\right)^{2}+\mu\sum_{k=1}^{r-1}\Psi_{e_{k}}\bar{\Psi}_{e_{k}}e^{b(e_{k},\varphi)},
\end{equation}
where  $\Psi_{e_{k}}$ is a parafermionic current corresponding to the simple root $e_{k}$ of $\mathfrak{sl}(r)$, $\bar{\Psi}_{e_{k}}$ its complex conjugate  and $\mathcal{L}_{\textrm{PF}}$ is a  formal Lagrangian for the parafermionic CFT. In these terms the integrable perturbation operator has the form $\Psi_{e_{0}}\bar{\Psi}_{e_{0}}e^{b(e_{0},\varphi)}$, where $e_{0}$  is the lowest root. Mathematically, one can define a system of local IM's as a set of local quantities in $\mathcal{U}(\mathcal{A}(r,p))$ which commute with all the fields
\begin{equation}
  \int_{\mathcal{C}}\Psi_{e_{k}}\bar{\Psi}_{e_{k}}e^{b(e_{k},\varphi)}dx,\qquad\text{for}\quad k=0,\dots,r-1.
\end{equation}
For $p=1$ this system is known to coincide with the quantum KdV type system with Lax operator of order $r$ \cite{Kupershmidt:1989bf}. This includes quantum KdV system for $r=2$ (related to the Virasoro algebra) and quantum Boussinesq system for $r=3$ (related to the $\mathrm{W}_{3}$ algebra). For generic values of the parameters $r$ and $p$ we call the corresponding integrable system quantum KdV system of type $(r,p)$ or $\textrm{qKdV}(r,p)$.

On the other hand the symmetry algebra of the coset \eqref{GKO} appeared recently in the context of AGT relation \cite{Alday:2009aq}. Namely, it was conjectured in \cite{Belavin:2011pp} that $U(r)$ instanton calculus on the quotient  $\mathbb{C}^{2}/\mathbb{Z}_{p}$ corresponds to the extended algebra
\begin{equation}\label{AGT-algebra}
   \widehat{\mathfrak{gl}}(p)_{r}\times\mathcal{A}(r,p).
\end{equation}
The original conjecture of \cite{Belavin:2011pp} was further checked in \cite{Nishioka:2011jk,Belavin:2011tb,Bonelli:2011jx,Alfimov:2011ju,Alfimov:2013cqa}. Mathematically, the results of \cite{Belavin:2011pp} predict the existence of a special basis  in the highest weight representation of the algebra $\widehat{\mathfrak{gl}}(p)_{r}\times\mathcal{A}(r,p)$ with remarkable property of factorization of matrix elements. In fact, as was emphasized in \cite{Belavin:2013fk,Belavin:2013kx}, there are different bases  corresponding to different compactifications of the space of instantons on  $\mathbb{C}^{2}/\mathbb{Z}_{p}$. For our purposes  the so called ``colored'' basis (see \cite{Belavin:2013kx} for details) is more suitable. This basis is an eigenbasis for the commutative algebra (Integrals of Motion) inside universal enveloping algebra of $\widehat{\mathfrak{gl}}(p)_{r}\times\mathcal{A}(r,p)$  with remarkably simple spectrum. For example, in the case of $r=1$ we get the spin Calogero-Sutherland integrable system \cite{Ha:1992zz}, whose eigenfunctions are known as Uglov polynomials \cite{Uglov:1997ia} (also known as $\mathfrak{gl}(p)$-Jack polynomials). In the case of $r>1$ the corresponding integrable system can be viewed as coupled system of $r$ copies of spin Calogero-Sutherland models. We call this system the generalized Calogero-Sutherland integrable system of the type $(r,p)$ or $\textrm{CS}(r,p)$.

In these notes we consider $p$-parametric integrable system which interpolates between $\textrm{CS}(r,p)$ and $\textrm{qKdV}(r,p)$. Namely, we state that there exists an  infinite  commutative family of  operators $\mathbf{I}_{k}\in\mathcal{U}(\widehat{\mathfrak{gl}}(p)_{r}\times\mathcal{A}(r,p))$ of spins $k=1,2,\dots$ depending on $p$ additional parameters $(q_{1},\dots,q_{p})$ which coincides with the $\textrm{CS}(r,p)$ system  for $(q_{1},\dots,q_{p})=(0,\dots,0)$ and degenerates to $\textrm{qKdV}(r,p)$ in the limit
\begin{equation}\label{local-limit}
  (q_{1},\dots,q_{p})\rightarrow (1,\dots,1).
\end{equation}
We call this integrable system quantum Intermediate Long Wave system of type $(r,p)$ or $\textrm{qILW}(r,p)$ for shortness. The reason for that is the same as above: for $p=1$ this system is known to be quantum  $\mathfrak{gl}(r)$ ILW  system. 

The main result of our paper is the claim that the common spectrum of $\mathbf{I}_{k}\in\mathcal{U}(\widehat{\mathfrak{gl}}(p)_{r}\times\mathcal{A}(r,p))$ is governed by Bethe equations. In order to write down these equations we introduce three functions
\begin{equation}\label{S-matrices}
  \mathbb{S}_{0}(x)=\frac{x+Q}{x-Q},\quad \mathbb{S}_{+}(x)=\frac{x-b}{x+b^{-1}},\quad \mathbb{S}_{-}(x)=\frac{x-b^{-1}}{x+b}.
\end{equation}
Then, for given $r$ and $p$ let $(N_{1},\dots,N_{p})$ be a set of non-negative integers and $\mathbb{A}_{k}(x)$ ($k=1,\dots,p$) be a set polynomials of degree $d_{k}$:
\begin{equation}\label{A-pols}
  \mathbb{A}_{k}(x)=\prod_{j=1}^{d_{k}}(x+iP_{j}^{(k)}),
\end{equation}
such that $\sum_{k=1}^{p}d_{k}=r$ and $\sum_{k=1}^{p}\sum_{j=1}^{d_{k}}P_{j}^{(k)}=0$. All these parameters correspond in a certain way to the representation data of $\widehat{\mathfrak{gl}}(p)_{r}\times\mathcal{A}(r,p)$. Namely, the numbers $(d_{1},\dots,d_{p})$ parameterize the highest weight of $\widehat{\mathfrak{gl}}(p)_{r}$, the continuous parameters $P_{j}^{(k)}$ correspond to the zero modes of the bosonic field $\varphi$ in \eqref{Lagrangian} and the set $(N_{1},\dots, N_{p})$ corresponds to the level. With each $N_{k}$ one associates a set of variables
\begin{equation*}
   x_{j}^{(k)},\quad j=1,\dots,N_{k},
\end{equation*}
where the cyclic symmetry is assumed $x_{j}^{(k)}=x_{j}^{(p+k)}$, $N_{k}=N_{p+k}$. Then our conjecture states:
\paragraph{Conjecture:} The spectrum of  $\textrm{qILW}(r,p)$ is governed by  the Bethe anzatz equations
\begin{equation}\label{BA}
   \frac{\mathbb{A}_{k}\bigl(x_{j}^{(k)}-\frac{Q}{2}\bigr)}{\mathbb{A}_{k}\bigl(x_{j}^{(k)}+\frac{Q}{2}\bigr)}
   \prod_{i=1}^{N_{k}}\mathbb{S}_{0}\bigl(x_{j}^{(k)}-x_{i}^{(k)}\bigr)
   \prod_{i=1}^{N_{k+1}}\mathbb{S}_{+}\bigl(x_{j}^{(k)}-x_{i}^{(k+1)}\bigr)
   \prod_{i=1}^{N_{k-1}}\mathbb{S}_{-}\bigl(x_{j}^{(k)}-x_{i}^{(k-1)}\bigr)=-q_{k},
\end{equation}
for $j=1,\dots, N_{k}$, $k=1,\dots, p$.

In order to make our conjecture more precise we need to construct IM's $\mathbf{I}_{k}$ explicitly and establish the relation between their eigenvalues and  roots of \eqref{BA}.  It is hard to do it in a full generality. This is why we consider only particular cases of our conjecture in next sections.  Namely, in section \ref{r} we consider the case of $p=1$ and generic $r$ which corresponds to the $\mathrm{W}_{r}$ algebras, in section \ref{p} we consider ``orthogonal'' direction: $r=1$ and generic $p$ corresponding to the free fermion theory $\widehat{\mathfrak{gl}}(p)_{1}$ and finally in section \ref{(2,2)} we consider a special case $p=r=2$ which corresponds to the $\textsf{NSR}$ algebra. Details of the generic case will be given in a separate publication.

Let us make a few remarks. First, as explained above, physically interesting situation corresponds to the limit  \eqref{local-limit}. In this limit the $\widehat{\mathfrak{gl}}(p)_{r}$ part of the symmetry algebra decouples and the system \eqref{BA} describes spectrum of IM's for the coset CFT \eqref{GKO} ($\textrm{qKdV}(r,p)$ system). However, for numerical purposes it is convenient to keep the parameters $q_{k}$ (twist parameters) away from criticality. Second, there is a relation between quantum integrable systems of ILW type and quantum cohomology \cite{Nekrasov:2009uh,2009arXiv0906.3587O,2009JAMS...22.1055M,Bonelli:2014iza,Nawata:2014nca}. In our approach we do not use this relation. Third, all our statements are of conjectural nature. They are based on explicit calculations for the lower levels. The derivation of Bethe anzatz equations \eqref{BA} from the first principles is an open problem. We discuss this problem as well as other interesting questions in concluding section \ref{concl}.
\section{The case $p=1$ and generic $r$:  W-algebras}\label{r}
In this case we have $\widehat{\mathfrak{gl}}(1)_{r}\times\mathcal{A}(r,1)=\mathrm{H}\times\mathrm{W}_{r}$. The algebra $\mathcal{A}(r,1)=\mathrm{W}_{r}$ is generated by holomorphic currents $T(z)$ and $W(z)$ of spins $2$ and $3$. Currents with higher spins from $4$ to $r$
appear in the OPE of $W(z)$. We fix $W(z)$ as usual: it is primary field of spin $3$  normalized as
\begin{equation}
    W(\xi)W(z)=\frac{c}{3(\xi-z)^{6}}+\dots
\end{equation}
with $c=(r-1)(1+r(r+1)Q^{2})$ (see \eqref{central-charge}). The commutation relations of the components of these currents are rather cumbersome. For $r=3$ they can be found in \cite{Fateev:1987vh}.  The Heisenberg algebra $\mathrm{H}$ is given by commutation relations (we set $a_{0}=0$ for convenience)
\begin{equation}
   [a_{m},a_{n}]=m\delta_{m,-n}.
\end{equation}
Working with $\mathrm{W}_{r}$ algebras it is convenient to use $\mathfrak{sl}(r)$ Lie algebra notations: $e_{k}$ are the simple roots of $\mathfrak{sl}(r)$, $\omega_{k}$ are the  fundamental weights, $h_{k}$ are the weights of fundamental representation and $\rho$ is the Weyl vector. We define the highest weight state $|\mathcal{P}\rangle$ where $\mathcal{P}$ is the $r-1$ component momentum as the state which is ``killed'' by the positive part of  the algebra $\textrm{H}\times\mathrm{W}_{r}$ and is an eigenstate for its Cartan subalgebra:
\begin{equation}
  a_{n}|\mathcal{P}\rangle=L_{n}|\mathcal{P}\rangle=W_{n}|\mathcal{P}\rangle=\dots=0\quad n>0,\quad
  L_{0}|\mathcal{P}\rangle=\Delta(P)|\mathcal{P}\rangle,\quad W_{0}|\mathcal{P}\rangle=w(\mathcal{P})|\mathcal{P}\rangle,\dots.
\end{equation} 
Here $L_{n},W_{n},\dots$ are the components of the currents $T(z),W(z),\dots$ and $(\Delta(\mathcal{P}),w(\mathcal{P}),\dots)$ are some Weyl invariant functions of $\mathcal{P}$. In particular, 
\begin{equation}
  \Delta(\mathcal{P})=\frac{1}{2}\left(\mathcal{Q}+i\mathcal{P},\mathcal{Q}-i\mathcal{P}\right)=\frac{r(r^{2}-1)Q^{2}}{24}+\frac{1}{2}\sum_{k=1}^{r}P_{k}^{2},\quad
  \text{here}\quad \mathcal{Q}=Q\rho\quad\text{and}\quad
  P_{k}=(\mathcal{P},h_{k}).
\end{equation}
The highest weight representation $\pi_{\mathcal{P}}$ of the algebra $\textrm{H}\oplus\mathrm{W}_{r}$ is formed from $|\mathcal{P}\rangle$ by creation operators. If the momentum $\mathcal{P}$ is generic then the corresponding representation is irreducible.  The number of states on  given level $N$ (the eigenvalue of the operator $L_{0}^{\text{total}}-\Delta$) is $\mathfrak{p}(r,N)$, where $\mathfrak{p}(r,N)$ is  the number of $r-$partitions of $N$.

We define the system of IM's $\mathbf{I}_{k}\in\mathcal{U}(\textrm{H}\oplus\mathrm{W}_{r})$ which depends on additional parameter $q_{1}=q$. First two IM's have the form
\begin{equation}\label{ILW-n}
    \begin{aligned}
       &\mathbf{I}_{1}=L_{0}+\sum_{k=1}^{\infty}a_{-k}a_{k}-\frac{c+1}{24},\\
       &\mathbf{I}_{2}=\sqrt{\frac{r(r^{2}-4)Q^{2}}{2}+2(r-2)}\,W_{0}+2\sum_{k\neq0}L_{-k}a_{k}+
       ir\sqrt{r}Q\sum_{k=1}^{\infty}k\frac{1+q^{k}}{1-q^{k}} a_{-k}a_{k}+
       \frac{1}{3}\sum_{i+j+k=0}a_{i}a_{j}a_{k}.
    \end{aligned}
\end{equation}
The remaining integrals $\mathbf{I}_{k}$  can be in principle determined from the condition of mutual commutativity with $\mathbf{I}_{1}$ and $\mathbf{I}_{2}$ (see \cite{Litvinov:2013zda} for expressions for higher IM's for $r=2$). Sometimes it is convenient to rewrite IM's as integrals of density functions. For example, we can write $\mathbf{I}_{2}$ as
\begin{equation}
  \mathbf{I}_{2}=\frac{1}{2\pi}\int_{0}^{2\pi}\left(\sqrt{\frac{r(r^{2}-4)Q^{2}}{2}+2(r-2)}\,W(x)+2T(x)J(x)+
  \frac{ ir\sqrt{r}Q}{2}J(x)\mathcal{D}J(x)+\frac{1}{3}J(x)^{3}\right)dx,
\end{equation}
where $J(x)$ is a $U(1)$ current $J(x)=\sum_{k}a_{k}e^{-ikx}$ and $\mathcal{D}$ is an operator whose Fourier symbol is $k\frac{1+q^{k}}{1-q^{k}}$.

We consider the spectral problem for the integrable system \eqref{ILW-n}. Our general conjecture \eqref{BA} applied to this case states that the spectrum of \eqref{ILW-n} on level $N$ is given by the Bethe anzatz equations 
\begin{equation}\label{BA-equations}
   \frac{\mathbb{A}\left(x_{i}-\frac{Q}{2}\right)}{\mathbb{A}\left(x_{i}+\frac{Q}{2}\right)}
   \prod_{j\neq i}\mathbb{S}(x_{i}-x_{j})=q,\quad\text{for}\quad i=1,\dots,N,
\end{equation}
where
\begin{equation*}
     \mathbb{S}(x)=\frac{(x-b)(x-b^{-1})(x+Q)}{(x+b)(x+b^{-1})(x-Q)},\quad
     \mathbb{A}(x)=\prod_{k=1}^{r}\left(x+iP_{k}\right).
\end{equation*}
Actually, we can formulate the statement more precisely and express the eigenvalues of the operators $\mathbf{I}_{k}$ in terms of Bethe roots. For example, the eigenvalues of the operator $\mathbf{I}_{2}$ are given by
\begin{equation}
   \mathbf{I}_{2}\sim\mathbf{I}_{2}^{\textrm{vac}}+2i\sqrt{r}\sum_{k=1}^{N}x_{k},
\end{equation}
where $\mathbf{I}_{2}^{\textrm{vac}}$ is an eigenvalue of the primary state $|\mathcal{P}\rangle$. Eigenvalues for other IM's are given by higher order power sum symmetric polynomials (more eigenvalues for the case $r=2$ can be found in \cite{Litvinov:2013zda}).

Let us comment on a local limit $q\rightarrow1$. In this limit the Heisenberg part of the symmetry algebra decouples and the system \eqref{BA-equations} describes the spectrum of local IM's in the $\mathrm{W}_{r}$ algebra or $\textrm{qKdV}(r,1)$ in our notations. First two IM's in this system have the form
\begin{equation}
   \mathbf{I}_{1}^{\text{local}}=L_{0}-\frac{c}{24},\qquad \mathbf{I}_{2}^{\text{local}}=W_{0}.
\end{equation}
On the other hand, the limit $q\rightarrow1$ of the integral $\mathbf{I}_{2}$ from \eqref{ILW-n} is formally singular.  In fact, the singularity is cancelled if one restricts the operator $\mathbf{I}_{2}$ to the purely $\mathrm{W}_{r}$ subspace (the set of states $|\lambda\rangle\in\pi_{\mathcal{P}}$: $a_{n}|\lambda\rangle=0$ for $n>0$).  On this subspace
\begin{equation}
   \mathbf{I}_{2}^{\text{local}}=\frac{1}{\sqrt{\frac{r(r^{2}-4)Q^{2}}{2}+2(r-2)}}\,\mathbf{I}_{2}.
\end{equation}
The same is true for higher order integrals $\mathbf{I}_{k}$. We note also that for $r=1$ nothing remains from the symmetry algebra in the local limit. Technically, it is equivalent to the statement that BA equations \eqref{BA-equations} have no solutions for $r=1$ and $q=1$. We note also that the spectrum of  $\mathrm{qKdV}(2,1)$ system was computed in \cite{Bazhanov:2004fk} using the so called ODE/IM correspondence. Our equations are different from those  in \cite{Bazhanov:2004fk} (see discussion in \cite{Litvinov:2013zda}).
\section{The case $r=1$ and generic $p$: $\widehat{\mathfrak{gl}}(p)_1$ algebra}\label{p}
In this case $\mathcal{A}(1,p)=0$. The remaining part of the symmetry algebra $\widehat{\mathfrak{gl}}(p)_1$ can be represented by $p$ complex fermions
\begin{equation}
  \psi^{a}(x)=\sum_{s}\psi^{a}_{s}e^{-isx},\quad \bar{\psi}^{a}(x)=\sum_{s}\bar{\psi}^{a}_{s}e^{-isx},\quad a=1,\dots, p,
\end{equation}
with anticommutation relations
\begin{equation}\label{fermions-comm-relat}
   \{\psi^{a}_{s},\psi^{b}_{r}\}=\{\bar{\psi}^{a}_{s},\bar{\psi}^{b}_{r}\}=0,\quad
   \{\psi^{a}_{s},\bar{\psi}^{b}_{r}\}=\delta^{a,b}\delta_{r,-s}.
\end{equation}
Here we consider only Neveu-Schwarz sector, i.e. $s,r\in\mathbb{Z}+\frac{1}{2}$. The highest weight  representation $\mathcal{F}$ of the fermionic algebra \eqref{fermions-comm-relat} is obtained in a usual way from the vacuum state $|0\rangle$ defined  by $\psi^{a}_{s}|0\rangle=\bar{\psi}^{a}_{s}|0\rangle=0$ for $s>0$. This representation naturally decomposes into the direct sum  $\mathcal{F}=\oplus\mathcal{F}_{l}$  where $\mathcal{F}_{l}$ is charge $l$ fermionic Fock module 
\begin{equation}
   \mathcal{F}_{l}=\textrm{Span}\Biggl\{\prod_{a=1}^{p}\psi^{a}_{-s_{1}^{a}}\dots\psi^{a}_{-s_{n_{a}}^{a}}
   \bar{\psi}^{a}_{-r_{1}^{a}}\dots\bar{\psi}^{a}_{-r_{m_{a}}^{a}}|0\rangle:\sum_{a=1}^{p}n_{a}-\sum_{a=1}^{p}m_{a}=l\Biggr\}.
\end{equation}
Each Fock module $\mathcal{F}_{l}$ forms an irreducible representation of $\widehat{\mathfrak{gl}}(p)_1$. They are all isomorphic to each other. For simplicity, we consider only $\mathcal{F}_{0}$.

According to our conjecture there exists an integrable system in $\mathcal{U}(\widehat{\mathfrak{gl}}(p)_{1})$ which depends on $p$ external parameters $q_{1},\dots,q_{p}$ whose spectrum is described by \eqref{BA}. It would be convenient to introduce the other parameters $q$ and $\alpha_{1},\dots,\alpha_{p-1}$ by
\begin{equation}
  q_{1}=\frac{q}{\alpha_{1}\dots \alpha_{p-1}},\quad q_{2}=\alpha_{1},\quad\dots\quad, q_{p}=\alpha_{p-1}.
\end{equation}
Then the first two Integrals of Motion of our integrable system are (here $::$ stands for the Wick ordering)
\begin{equation}\label{I-2}
\begin{aligned}
  &\mathbf{I}_{1}=\frac{1}{2\pi}\int_{0}^{2\pi}\hspace*{-8pt}:\bar{\psi}^{a}\partial\psi^{a}:dx,\\
  &\mathbf{I}_{2}=\frac{1}{2\pi}\int_{0}^{2\pi}\hspace*{-8pt}:\hspace*{-2pt}\left(\frac{i(b-b^{-1})}{2}\bar{\psi}^{a}\partial^{2}\psi^{a}
  +\frac{iQ}{2p}:\left(\partial\psi^{a}\bar{\psi}^{a}+\partial\bar{\psi}^{a}\psi^{a}\right):\mathbf{H}\,:\bar{\psi}^{b}\psi^{b}:
  +\frac{iQ}{2p}:\bar{\psi}^{a}\psi^{b}:\mathbf{D}^{ab}:\bar{\psi}^{b}\psi^{a}:\right)\hspace*{-2pt}:dx,
\end{aligned}
\end{equation}
where the Fourier symbols of the operators $\mathbf{H}$ and $\mathbf{D}^{ab}$ are
\begin{equation}
  \mathbf{H}(k)=\textrm{sgn}(k),\qquad
  \mathbf{D}^{ab}(k)=k\,\frac{1+\alpha_{a}\dots \alpha_{b-1}\,q^{k}}{1-\alpha_{a}\dots \alpha_{b-1}\,q^{k}}\quad\text{for}\quad a\leq b\quad\text{and}\quad
  \mathbf{D}^{ba}(k)=\mathbf{D}^{ab}(-k),
\end{equation}
and the sum over the repeated indexes is assumed. In principle, one can find higher IM's from the condition of mutual commutativity with $\mathbf{I}_{1}$ and $\mathbf{I}_{2}$, but their explicit form is more cumbersome.

The space $\mathcal{F}_{0}$ has natural gradings: the level $N$ and the fermionic numbers $\nu_{a}$ defined by
\begin{equation}
   N=\sum_{a=1}^{p}\bigl(\sum_{i=1}^{n_{a}}s_{i}^{a}+\sum_{j=1}^{m_{a}}r_{j}^{a}\bigr),\quad \nu_{a}=n_{a}-m_{a}.
\end{equation}
Hence this space decomposes into the direct sum $\mathcal{F}_{0}=\oplus\mathcal{F}_{0}^{(N,\vec{\nu})}$ where
\begin{equation}
   \mathcal{F}_{0}^{(N,\vec{\nu})}=\textrm{Span}\Biggl\{\prod_{a=1}^{p}\psi^{a}_{-s_{1}^{a}}\dots\psi^{a}_{-s_{n_{a}}^{a}}
   \bar{\psi}^{a}_{-r_{1}^{a}}\dots\bar{\psi}^{a}_{-r_{m_{a}}^{a}}|0\rangle:\sum_{a=1}^{p}\bigl(\sum_{i=1}^{n_{a}}s_{i}^{a}+\sum_{j=1}^{m_{a}}r_{j}^{a}\bigr)=N,\,n_{a}-m_{a}=\nu_{a}\Biggr\}.
\end{equation}
Clearly, the integrals $\mathbf{I}_{k}$ conserve the gradings. Therefore, our spectral problem decomposes into the collection of finite dimensional spectral problems  on $\mathcal{F}_{0}^{(N,\vec{\nu})}$. We claim that the spectrum on $\mathcal{F}_{0}^{(N,\vec{\nu})}$  is governed by the BA equations \eqref{BA} with
\begin{equation}
   A_{1}(x)=x,\quad A_{2}(x)=\dots=A_{p}(x)=1,\quad N_{k}=N+\sum_{a=1}^{k-1}\nu_{a}.
\end{equation}
In particular, the spectrum of the integral $\mathbf{I}_{2}$ is given by the sum 
\begin{equation}
  \mathbf{I}_{2}\sim\frac{2i}{p}\sum_{j=1}^{N_{1}}x_{j}^{(1)}.
\end{equation}

We note that in the limit $(q_{1},\dots,q_{p})\rightarrow(0,\dots,0)$ the integral $\mathbf{I}_{2}$ coincides with the Hamiltonian of the  spin Calogero model \cite{Ha:1992zz} written in a second quantized form (see \cite{Aniceto:2006rr}). One can easily see that the spectrum  is degenerate in this limit. The reason for this degeneracy is well known \cite{Bernard:1993uz}: the Integrals of Motion of the spin Calogero model commute with the Yangian $Y(\mathfrak{gl}(p))$. We were unable to find if there is a part of this symmetry which survives the perturbation (switching-on of the parameters $q_{k}$).
\section{The case $r=p=2$:  quantum SUSY KdV system}\label{(2,2)}
Here, for simplicity, we consider only the local case $q_{1}=q_{2}=1$. According to our conjecture $\widehat{\mathfrak{gl}}(2)_{2}$ part of the algebra \eqref{AGT-algebra} decouples in this limit. Then we use the fact that $\mathcal{A}(2,2)=\textsf{NSR}$, where $\textsf{NSR}$ is a superconformal algebra (Neveu-Schwarz-Ramond algebra). This algebra is generated by currents $T(z)$ and $G(z)$ of spins $2$ and $3/2$. Their components $L_{n}$'s and $G_{r}$'s satisfy the relations
\begin{equation}\label{NS-comm-relat}
\begin{gathered}
[L_{n},L_{m}]=(n-m)L_{n+m}+\frac{c}{8}(n^{3}-n)\delta_{n+m,0},\\
\{G_{r},G_{s}\}= 2L_{r+s}+\frac{1}{2}c\,(r^{2}-\frac{1}{4})\delta_{r+s,0},\quad
[L_{n},G_{r}]=\left(\frac{1}{2}n-r\right)G_{n+r},
\end{gathered}
\end{equation}
where according to \eqref{central-charge}: $c=1+2Q^{2}$. Below we will  consider only Neveu-Schwarz  sector, so the indexes $r$ and $s$ in \eqref{NS-comm-relat} take half-integer values. The highest weight representation $\pi_{P}$ of \eqref{NS-comm-relat}  is defined by the vacuum state $|P\rangle$
\begin{equation}\label{NS-vac}
    L_{n}|P\rangle=G_{r}|P\rangle=0\quad\text{for}\quad n,r>0,\qquad L_{0}|P\rangle=\Delta(P)|P\rangle,
\end{equation}
where $\Delta(P)=\frac{1}{2}\left(\frac{Q^{2}}{4}+P^{2}\right)$, and consists of the vectors of the form
\begin{equation}\label{NSR-states}
  L_{-n_{1}}\dots L_{-n_{k}}G_{-s_{1}}\dots G_{-s_{l}}|P\rangle,\quad\text{where}\quad n_{1}\geq n_{2}\dots,\;
  s_{1}>s_{2}>\dots
\end{equation}

The integrable system which appears in this case coincides with the quantum SUSY KdV system \cite{Mathieu:1989nr} whose local Integrals of Motion have the form\footnote{Here all the fields are assumed to be analytically regularized. See for example \cite{Bazhanov:1994ft}.}
\begin{equation}\label{NSR-KdV}
      \mathbf{I}_{1}=\frac{1}{2\pi}\int_{0}^{2\pi}Tdx,\qquad
      \mathbf{I}_{3}=\frac{1}{2\pi}\int_{0}^{2\pi}\left(T^{2}+\frac{i}{4}GG'\right)dx,\qquad \dots\dots
\end{equation}
The spectral problem for the quantum SUSY KdV system was considered in \cite{Kulish:2004ap,Kulish:2005qc}. We suggest a different approach to the same problem. For us it will be important, that there is also a set of non-local Integrals of Motion of integer spin which commute with the local ones. First of them has the form
\begin{equation}
      \tilde{\mathbf{I}}_{1}=\frac{1}{2\pi}\int_{0}^{2\pi}\left(G\partial^{-1}G\right)dx.
\end{equation}
The expressions for higher non-local integrals are known only in the classical limit \cite{Andrea:2008ez}, but there is no doubts that one can find their quantum analogs.

We claim that the spectrum of quantum SUSY KdV system in Neveu-Schwarz  sector is governed by Bethe Anzatz equations \eqref{BA} with  $p=r=2$, $q_{1}=q_{2}=1$ and
\begin{equation}
   A_{1}(x)=x^{2}+P^{2},\qquad A_{2}(x)=1.
\end{equation}
Exact expressions for the eigenvalues depend on the value of the level $N=\sum_{k}n_{k}+\sum_{r}s_{r}$ in \eqref{NSR-states} which can be either integer or half-integer. Namely, one has
\begin{itemize}
\item For integer $N$ we take $N_{1}=N_{2}=N$ in \eqref{BA}. Then the eigenvalues of the operator $\mathbf{I}_{3}+\frac{9}{16}i\,\tilde{\mathbf{I}}_{1}$ are given by
\begin{equation}
   \mathbf{I}_{3}+\frac{9}{16}i\,\tilde{\mathbf{I}}_{1}\sim\mathbf{I}_{3}^{\textrm{vac}}(\Delta+N)+4N\Delta+2N(N-2)+6\sum_{k=1}^{N}(x_{k}^{(1)})^{2}.
\end{equation}
\item For half-integer $N$ we take $N_{1}=N+\frac{1}{2}$, $N_{2}=N-\frac{1}{2}$ in \eqref{BA}, then
\begin{equation}
   \mathbf{I}_{3}-\frac{15}{16}i\,\tilde{\mathbf{I}}_{1}\sim\mathbf{I}_{3}^{\textrm{vac}}\left(\Delta+N\right)+4\left(N+\frac{3}{2}\right)\Delta+
   2\left(N+\frac{3}{2}\right)\left(N-\frac{1}{2}\right)+6\sum_{k=1}^{N+\frac{1}{2}}(x_{k}^{(1)})^{2},
\end{equation}
\end{itemize}
here $\mathbf{I}_{3}^{\textrm{vac}}(\Delta)$ is the eigenvalue of the primary state. We note here the phenomenon which was not present in the cases considered before. Namely, only special combinations of local and non-local IM's have eigenvalues which depend polynomially on Bethe roots. On lower levels we found that the eigenvalues of the local  integral $\mathbf{I}_{3}$ are some rational symmetric functions of $x_{k}^{(1)}$, but their general form is not known.
\section{Conclusions}\label{concl}
In these notes we studied the spectrum of the local Integrals of Motion in the coset CFT \eqref{GKO}. In fact, we noticed that it is quite natural to extend the coset algebra $\mathcal{A}(r,p)$ by $\widehat{\mathfrak{gl}}(p)_{r}$. We proposed that there exists the system of IM's in the universal enveloping of the extended algebra which depends on $p$ external parameters. We called this system $\textrm{qILW}(r,p)$ system and claimed that its spectrum is given by the Bethe anzatz equations \eqref{BA}. We have verified our conjecture for particular values of $r$  and $p$. Let us make a few comments which seem to be important. 
\begin{enumerate}
\item It is now clear that the integrable system considered in this paper follows from certain $R$-matrix with parameters $q_{k}$ being twist parameters. It can be thought as an $R-$matrix for the Yangian $Y(\widehat{\mathfrak{gl}}(p))$ \cite{Maulik:2012wi}. What is not known is the chain of arguments leading from this $R-$matrix to the Bethe equations \eqref{BA}. In other words we do not know  how to develop the Bethe anzatz (algebraic, functional or whatever else)  in this case.
\item As we explained in the introduction the chiral algebra description of the coset CFT \eqref{GKO} is too hard to analyze due to the presence of the currents with non-abelian braiding. It is believed that there exist local symmetry description of the same CFT. Such a description was found in \cite{Fateev:1996ea,Feigin:2001yq} for $r=2$. The corresponding chiral algebra is generated by $W$ current of spin $4$ and can be realized in terms of three bosonic fields. One of the advantages of this description is that it is continuous in both parameters of the coset \eqref{GKO}. In our approach one of the parameters of the coset \eqref{GKO} is necessarily an integer (we take $p$ to be  integer and $n$ to be continuous). It would be interesting to find a way to generalize the Bethe anzatz equations \eqref{BA} for non-integer values of $p$. We note that the spectrum of local IM's in three-field theory (also known as Fateev model) was computed in \cite{Lukyanov:2013wra,Bazhanov:2013cua} using ODE/IM correspondence. The relation between ODE/IM approach and ours is not clear.
\end{enumerate}
\section*{Acknowledgements}
A.~L. thanks Sergei Lukyanov, Pierre Mathieu, Alexei Oblomkov, Andrei Zayakin and especially Mikhail Bershtein for interesting discussions.  This research was done with the support of the Russian Scientific Foundation under the grant 14-12-01383.

\bibliographystyle{MyStyle} 
\bibliography{MyBib}
\end{document}